\documentclass{llncs}
 
\usepackage{amssymb}
\usepackage{mathrsfs}
\usepackage{graphicx}

\def\db{D}

\def\tinya{\mbox{\tiny ${\cal A}$}}
\def\tinyb{\mbox{\tiny ${\cal B}$}}

\def\tinyel{\mbox{\tiny ${\cal L}$}}

\def\lsem{\lbrack}
\def\rsem{\rbrack}

\begin{document}

\title{Provenance for Regular Path Queries}

\author{
G\"{o}sta Grahne \inst{1} 
\and        
Alex Thomo \inst{2}
}

\institute{
Concordia University, Montreal, Canada,
\email{grahne@cs.concordia.ca}
\and
University of Victoria,
Victoria, Canada, 
\email{thomo@cs.uvic.ca}
}

\date{}

\maketitle


\section{Introduction}

It has been recognized that
the result of a database query should be annotated
with {\em provenance}, i.e.\ information about
how, why, where, with what level of certainty
or security clearance, etc a particular fact of
the query was derived. The seminal paper by
Green, Tannen and Karvounarakis 
\cite{DBLP:conf/pods/GreenKT07}
convincingly showed that all major forms of
provenance can be uniformly captured within
the algebraic framework of {\em semirings}.
Green et al.\ show that a suitably 
semiring-annotated
positive (negation-free) relational algebra
and datalog
can capture the provenance of query results.
Furthermore, the various data base semirings
form a partial order where coarser ("smaller")
semirings
can be obtained as homomorphic images of
semirings with a finer grain of information.
Green et al.\ also show that the
annotated positive relational algebra
and datalog form congruences
within their semiring hierarchy.

{\em Regular path queries} (RPQs) 
is the ubiquitous mechanism for 
querying graph databases
\cite{DBLP:journals/jcss/CalvaneseGLV02}. 
RPQs are in essence regular expressions 
over the edge symbols. 
The answer to an RPQ on a given graph database 
is the set of pairs of objects $(a,b)$, 
which are connected by paths 
spelling words in the language of the regular path query
An annotated pair in the answer would naturally
contain the set of words that spell paths between
$a$ and $b$. However, a finer grain of provenance
can be obtained by annotating the words with the
intermediate vertices of each path spelling the word.

Since graph databases have their roots in
automata theory, and automata have their roots
in the algebraic theory of
semiring-automata \cite{DBLP:reference/hfl/Kuich97},
an investigation into how the
provenance algebra of Green et al.\ can be
paired with the algebra of semiring-automata
is called for.
The paper at hand represents a first step in this
direction.

\section{Databases and Regular Path Queries}
 \label{background}

We consider a database
to be an edge-labeled graph.
Intuitively,
the nodes of the database graph represent objects
and the edges represent relationships 
between the objects. 
The edge labels are drawn from a finite alphabet $\Delta$.
Elements of $\Delta$ will be denoted $r, s, \ldots$.
As usual, $\Delta^*$ denotes the set of all
finite words over $\Delta$. Words will be denoted by
$u,w,\ldots$.
We also assume that we have a universe of objects,
and objects will be denoted
$a, b, c\ldots$.

Associated with each edge is a weight expressing the ``strength''
of the edge. 
Such a ``strength'' can be multiplicity, cost, distance, etc, 
and is expressed by an element of some semiring 
$\mathscr{R} = (R, \oplus,\otimes, \mathbf{0}, \mathbf{1})$
where
\begin{enumerate}
\item $(R,\oplus,\mathbf{0})$ is a commutative monoid with $0$ as the identity 
	element for $\oplus$.
\item $(R,\otimes,\mathbf{1})$ is a monoid with $1$ as the identity 
	element for $\otimes$.
\item $\otimes$ distributes over $\oplus$: for all $x,y,z \in R$,
	\begin{eqnarray*}
	(x \oplus y) \otimes z &=& 
					(x \otimes z) \oplus (y \otimes z) \\
	z \otimes (x \oplus y) &=& 
					(z\otimes x  ) \oplus (z\otimes y).	
	\end{eqnarray*}
\item $\mathbf{0}$ is an anihilator for $\otimes$: $\forall x \in R$, 
	$x \otimes \mathbf{0} = \mathbf{0} \otimes x = \mathbf{0}$.
\end{enumerate}
For simplicity we will blur the distinction between $\mathscr{R}$ and $R$, 
and will only use $\mathscr{R}$ in our development.

In this paper, we will in addition require for semirings
to have a {\em total} order $\preceq$.
If $x\preceq y$, we say that $x$ is {\em better} than $y$. 
``Better'' will have a clear meaning depending on the context. 
 

Now, a {\em database} $\db$  
is formally a graph $(V,E)$, 
where $V$ is a finite set of objects and
$E \subseteq V \times \Delta \times \mathscr{R} \times V$
is a set of directed edges labeled with symbols from $\Delta$ 
and weighted by elements of $\mathscr{R}$.

We concatenate the labels along the edges of a path into words in $\Delta^*$.
Also, we aggregate the weights along the edges of a path by using the 
multiplication operator $\otimes$. 
Formally, let 
$\pi = (a_1,r_1,x_1,a_2),\ldots,(a_n,r_n,x_n,a_{n+1})$ 
be a path in $\db$. 
We define 
the {\em start}, the {\em end}, 
the {\em label}, and the {\em weight} of $\pi$ to be 
\begin{eqnarray*}
\alpha(\pi) &=& a_1 \\
\beta(\pi) &=& a_{n+1} \\
\lambda(\pi) &=& r_1\cdot \ldots \cdot r_n \in \Delta^* \\ 
\kappa(\pi) &=& x_1\otimes \ldots \otimes r_n \in \mathscr{R}
\end{eqnarray*}
respectively. 

A {\em regular path query} (RPQ) is a regular language over $\Delta$.
For the ease of notation, we will blur the distinction between
regular languages and regular expressions that represent them.
Let $Q$ be an RPQ and $\db = (V,E)$ a database.
Now let $a$ and $b$ be two objects in $\db$, and $w\in\Delta^*$. 
We define
\begin{eqnarray*}
\Pi_{\mbox{\tiny w,\db}}(a,b) &=& 
  \{\pi \mbox{ in } \db: \alpha(\pi)=a, \beta(\pi)=b, \lambda(\pi)=w\} \\
\Pi_{\mbox{\tiny Q,\db}}(a,b) &=& \bigcup_{w\in Q} \Pi_{\mbox{\tiny w,\db}}(a,b).
\end{eqnarray*}
Then, the {\em answer} to $Q$ on $\db$ 
is defined~as
\begin{eqnarray*}
 \mbox{\sl Ans}(Q,\db) &=& 
 		\{[(a,b),x] \in (V\times V) \times \mathscr{R} : \\
 			& &	\;\;\Pi_{\mbox{\tiny Q,\db}}(a,b)\neq\emptyset \mbox{ and }  \\
 			& &	\;\;x = \oplus\;\{\kappa(\pi) : \pi\in\Pi_{\mbox{\tiny Q,\db}}(a,b)\}\}.
\end{eqnarray*}
If $[(a,b),x] \in \mbox{\sl Ans}(Q,\db)$, we say that 
$(a,b)$ is an answer of $Q$ on $\db$ with weight~$x$.

\medskip
\noindent
Let $w\in Q$. Suppose $\Pi_{\mbox{\tiny w,\db}}(a,b) \neq \emptyset$. 
Clearly, $(a,b)$ is an answer of $Q$ on $\db$ with some weight $x$.   
We say $w$ is the {\em basis} of a ``reason'' for $(a,b)$ to be such an answer.
This basis has obviously a weight (or strength) coming with it, namely
$y = \oplus\Pi_{\mbox{\tiny w,\db}}(a,b)$.
We say that $(w,y)$ is a {\em reason} for $(a,b)$ to be an answer
of $Q$ on $\db$. 
In general, there can be many such reasons. 
We denote by 
$$\Xi_{\mbox{\tiny Q,\db}}(a,b)$$
the set of reasons 
for $(a,b)$ to be an answer of $Q$ on $\db$.
It can be seen that
$x = \oplus\;\{y : (w,y)\in \Xi_{\mbox{\tiny Q,\db}}(a,b)\}$. 

In the rest of the paper we will be interested in 
determining whether a pair $(a,b)$ has ``the same or stronger'' reasons 
than another pair $(c,d)$ to be in the answer of $Q$ on $D$.

Evidently, 
$\Xi_{\mbox{\tiny Q,\db}}(a,b) \subseteq \Delta^* \times \mathscr{R}$, 
but we have a stronger property for 
$\Xi_{\mbox{\tiny Q,\db}}(a,b)$. 
It is a partial function from $\Delta^*$ to $\mathscr{R}$.
We complete 
$\Xi_{\mbox{\tiny Q,\db}}(a,b)$ to be a function 
by adding $\bf 0$-weighted reasons. 

\medskip
\noindent
An $\mathscr{R}$-annotated language (AL) $L$ over $\Delta$ is a function
$$
L : \Delta^* \rightarrow \mathscr{R}.
$$
Frequently, we will write $(w,x)\in L$ instead of
$L(w) = x$.
From the above discussion, 
$\Xi_{\mbox{\tiny Q,\db}}(a,b)$ is such an AL. 

Given two $\mathscr{R}$-ALs $L_1$ and $L_2$, we say 
that $L_1$ is {\em contained} in $L_2$ iff 
$(w,x)\in L_1$ implies 
$(w,y)\in L_2$ and $x\preceq y$.

Now, we say that 
$\Xi_{\mbox{\tiny Q,\db}}(a,b)$ is {\em the same or stronger} than 
$\Xi_{\mbox{\tiny Q,\db}}(c,d)$, 
iff, 
$$\Xi_{\mbox{\tiny Q,\db}}(a,b) \preceq \Xi_{\mbox{\tiny Q,\db}}(c,d).$$

It might seem strange to use $\preceq$ to say 
``stronger'', but we are motivated 
by the notion of distance in real life. 
The shorter this distance, the stronger the relationship between 
two objects (or subjects) is. 
 
If
$\Xi_{\mbox{\tiny Q,\db}}(a,b) \preceq \Xi_{\mbox{\tiny Q,\db}}(c,d)$ and
$\Xi_{\mbox{\tiny Q,\db}}(c,d) \preceq \Xi_{\mbox{\tiny Q,\db}}(a,b)$, 
we say that $(a,b)$ and $(c,d)$ are in the answer of $Q$ on $D$
for {\em exactly the same reasons} and write
$$
\Xi_{\mbox{\tiny Q,\db}}(a,b) = \Xi_{\mbox{\tiny Q,\db}}(c,d).
$$

\section{Computing Reason Languages}
\label{answering}

An {\em annotated automaton} $\cal A$ is a quintuple
$(P,\Delta,\mathscr{R},\tau,p_0,F)$,
where $\tau$ is a subset of
$\mbox{$P \times \Delta \times \mathscr{R} \times P$}$.
Each annotated automaton $\cal A$ defines
an AL, 
denoted by $\lsem {\cal A} \rsem$ 
and defined by
\begin{eqnarray*}
\lsem {\cal A}\rsem = \{(w,x)\in\Delta^*\times{R}  & : & \\
										& & \!\!\!\!\!\!\!\!\!\!\!\!\!\!\!\!\!\!\!\!\!\!\!\!
										w=r_1r_2\ldots r_n, 
 x = \oplus \; \{ \otimes_{i=1}^n x_i : 
(p_{i-1},r_i,x_i,p_i)\in\tau,p_n\in F\} \}.
\end{eqnarray*}  

An AL $L$ is a {\em regular annotated language} (RAL),
if $L = \lsem A\rsem$, for some semiring automaton $\cal A$.

Given an RPQ $Q$, a database $\db$, and a pair $(a,b)$ of objects in $\db$, 
it turns out that 
the reason language $\Xi_{\mbox{\tiny Q,\db}}(a,b)$ is RAL. 

An annotated automaton for $\Xi_{\mbox{\tiny Q,\db}}(a,b)$
is constructed by computing 
a ``lazy'' Cartesian product of a (classical) automaton $\cal Q$ for $Q$ with 
database $\db$. 
For this we proceed by creating state-object pairs from 
the query automaton and the database. 
Starting from object $a$ in $\db$, 
we first create the pair $(p_0, a)$, 
where $p_0$ is the initial state in ${\cal Q}$. 
We then create all the pairs $(p,b)$ such that 
there exist a transition $t$ from $p_0$ to $p$ in ${\cal Q}$, 
and an edge $e$ from $a$ to $b$ in $\db$, and the 
labels of $t$ and $e$ match.
The weight of this edge is set to be the weight of edge $e$ in $\db$.

In the same way, we continue to create new pairs from existing ones,
until we are not anymore able to do so.
In essence, what is happening is a lazy construction of 
a Cartesian product graph of $\cal Q$ and $\db$. 
Of course, only a small (hopefully) part of the Cartesian product 
is really contructed depending on the selectivity of the query. 
The implicit assumption is that this part of the 
Cartesian product fits in main memory and each object is not accessed
more than once in secondary storage.  

Let us denote by ${\cal C}_{\mbox{\tiny Q,\db}}(a,b)$ the above Cartesian product. 
We can consider ${\cal C}_{\mbox{\tiny Q,\db}}(a,b)$ to be a weighted automaton 
with initial state $(a,p_0)$ and set of final states
$\{(p,b) : p\in F_{\mbox{\tiny {\cal Q}}}\}$, 
where $F_{\mbox{\tiny {\cal Q}}}$ 
is the set of final states of $\cal Q$.
It is easy to see that 
$$
\Xi_{\mbox{\tiny Q,\db}}(a,b) = \lsem {\cal C}_{\mbox{\tiny Q,\db}}(a,b) \rsem.
$$

\section{Some Useful Semirings}
\label{semirings}

We will consider the following semirings in this paper. 
\begin{description}
\item[{\em boolean}] 
		$\mathscr{B} = (\{T,F\},\vee,\wedge,F,T)$
\item[{\em tropical}] 
		$\mathscr{T} = (\mathbb{N}\cup\{\infty\}, \mbox{\em min}, +, \infty, 0)$
\item[{\em fuzzy}]
		$\mathscr{F} = 
(\mathbb{N}\cup\{\infty\},  
\mbox{\em min}, 
\mbox{\em max}, 
\infty, 0)$
\item[{\em multiplicity}]
$\mathscr{N} = (\mathbb{N}, +, \cdot, 0, 1)$.
\end{description}
$T$ and $F$ stand for ``true'' and ``false'' respectively, 
and $\vee$, $\wedge$ are the usual ``and'' and ``or'' Boolean operators. 
On the other hand, 
{\em min}, {\em max}, $+$, and $\cdot$ are the usual 
operators for integers.

It is easy to see that a Boolean annotated automaton
${\cal A} = (P,\Delta,\mathscr{B},\tau,p_0,F)$ is indeed
an ``ordinary'' finite state automaton 
$(P,\Delta,\tau,p_0,F)$, 
and a RAL over $\mathscr{B}$ is a an ``ordinary'' regular language over $\Delta$.
In this case it can be seen that 
$$
\Xi_{\mbox{\tiny Q,\db}}(a,b) \preceq \Xi_{\mbox{\tiny Q,\db}}(c,d)
\Leftrightarrow
\Xi_{\mbox{\tiny Q,\db}}(a,b) \supseteq \Xi_{\mbox{\tiny Q,\db}}(c,d).
$$
Since the containment of regular languages is decidable, we have that the provenance problem is decidable in the case of semiring $\mathscr{B}$.

For semiring $\mathscr{F}$ we show later that the problem is decidable. 

On the other hand, for semirings $\mathscr{T}$ and $\mathscr{N}$ the problem is 
unfortunately undecidable. 
For these results we refer to \cite{Krob92} and \cite{Eil76}, respectively.
\cite{Krob92} shows an even stronger result that 
the problem of RAL equivalence, 
which is $L_1\preceq L_2$ and $L_2\preceq L_1$ at the same time, 
is undecidable. 
On the other, it is interesting to note that for the case 
of $\mathscr{N}$, only the containment problem is undecidable, 
whereas the equivalence is in fact decidable in polynomial time 
via a reduction to a linear algebra problem~\cite{Eil76}.

\section{Spheres and Stripes}
\label{spheres}

Let $L$ be an annotated language over a semiring $\mathscr{R}$.
We have
\begin{definition} 
Let $x\in \mathscr{R}$.
\begin{enumerate}
\item 
The {\em $x$-inner sphere} of $L$ is
$$
L^{x} = \{(w,y) \in \Delta^* \times \mathscr{R}: (w,y) \in L 
														\mbox{ and } y \preceq x\}.
$$
\item 
The {\em $x$-outer sphere} of $L$ is
$$
L^{\breve{x}} = \{(w,y) \in \Delta^* \times \mathscr{R}: (w,y) \in L 
														\mbox{ and } x \preceq y\}.
$$
\item 
The {\em $x$-stripe} of $L$ is
$$
L^{\dot{x}} = \{(w,y) \in \Delta^* \times \mathscr{R}: (w,y) \in L 
														\mbox{ and } y = x\}.																										  
$$
\end{enumerate}
\end{definition}

We now give the following characterization theorem~\cite{GTW08}.
\begin{theorem} \label{containment}
Let $L_1$ and $L_2$ be two annotated languages over a discrete semiring~$\mathscr{R}$.
Then, $L_1 \preceq L_2$, if and only if, 
\begin{enumerate}
	\item $\lfloor L_1 \rfloor \subseteq \lfloor L_2 \rfloor$,
	\item $\lfloor L_2^{\dot{x}} \rfloor \cap \lfloor L_1 \rfloor \subseteq \lfloor L_1^{\breve{x}} \rfloor$,
	for each element $x$ of~$\mathscr{R}$. 
\end{enumerate}
\end{theorem}
{\em Proof.}
{\em If.}
Let $(w,x) \in L_1$. 
By condition (1), $w \in \lfloor L_2 \rfloor$, 
and thus, there exists $y$ in $\mathscr{R}$, 
such that $(w,y) \in L_2$.
Now, we want to show that $y\preceq x$. 
For this, observe that 
$w \in \lfloor L_2^{\dot{y}} \rfloor$ and since also $w \in \lfloor L_1 \rfloor$, 
we have $w \in \lfloor L_2^{\dot{y}} \rfloor \cap \lfloor L_1 \rfloor$.
By condition~(2),  
$\lfloor L_2^{\dot{y}} \rfloor \cap \lfloor L_1 \rfloor \subseteq \lfloor L_1^{\breve{y}} \rfloor$, 
i.e. 
$w \in \lfloor L_1^{\breve{y}} \rfloor$. 
The latter means that
$(w,x) \in L_1^{\breve{y}}$, i.e. $y\preceq x$.

{\em Only.} 
If $L_1 \sqsubseteq_{\tinyel} L_2$, then, clearly, condition (1) directly follows.
Now, let $w \in \lfloor L_2^{\dot{y}} \rfloor \cap \lfloor L_1 \rfloor$, for some $y$ in $\mathscr{R}$. 
From this, we have that $(w,y) \in L_2$ and $(w,x) \in L_1$ for some $x$ in $\mathscr{R}$. 
By the fact that $L_1 \preceq L_2$, 
$y \preceq x$. 
Thus, $(w,x) \in L_1^{\breve{y}}$, which in turn means that 
$w \in \lfloor L_1^{\breve{y}} \rfloor$. 
Since $y$ was arbitrary, we have that condition (2) is satisfied as well. 
\qed
\bigskip

Observe that 
that conditions (1) and (2) of Theorem~\ref{containment}
are about containment checks of pure 
languages that we obtain if we ignore 
the weight of the words in the corresponding annotated languages.  
These containments are decidable when $L_1$ and $L_2$ are RALs.

We have the following useful equalities
\begin{eqnarray} 
\lfloor L^{\breve{x}} \rfloor &=& (\lfloor L \rfloor \setminus \lfloor L^x \rfloor) \cup \lfloor L^{\dot{x}}\rfloor 
  \\
\lfloor L^{x} \rfloor &=& (\lfloor L \rfloor \setminus \lfloor L^{\breve{x}} \rfloor) \cup \lfloor L^{\dot{x}}\rfloor.
\end{eqnarray}

We also define here the notion of ``discrete'' semirings.
\begin{definition}
A semiring ${\mathscr{R}} = (R, \oplus,\otimes, \mathbf{0}, \mathbf{1})$ 
is said to be {\em discrete} 
iff for each
$x\not= \mathbf{0}$ in $R$ there exists $y$ in $R$, 
such that 
\begin{enumerate}
\item $x \prec y$, and 
\item there does not exist $z$ in $R$, such that $x \prec z \prec y$. 
\end{enumerate}
$y$ is called the {\em next element after $x$}, 
whereas 
$x$ is called the {\em previous element before $y$}. 
\end{definition}
Observe that all the semirings we list in Section~\ref{semirings} are discrete.

For all the discrete semirings,
we can compute $L^{\dot{x}}$ by 
computing 
$\lfloor L^x \rfloor \setminus \lfloor L^u \rfloor$
or 
$\lfloor L^{\breve{u}} \rfloor \setminus \lfloor L^{\breve{x}} \rfloor$
where $u$ is the previous element before~$x$.
[Initially, $\lfloor L^{\dot{\bf{1}}} \rfloor = \lfloor L^{\bf{1}} \rfloor$.]
For such semirings then, in order to decide $L_1 \preceq L_2$
based on Theorem~\ref{containment}, we need to be able to compute 
either inner or outer spheres.  

Nevertheless, 
Theorem~\ref{containment} does not necessarily give a decision procedure 
for $L_1 \preceq L_2$ when semirings $\mathscr{T}$ and $\mathscr{N}$ are considered, even if $L_1$ and $L_2$ are RALs. 
This is because the number of inner (and outer) spheres might be infinite
for these semantics.  

Interestingly, Theorem~\ref{containment} gives an effective 
procedure for deciding $L_1 \preceq L_2$ when the fuzzy semiring $\mathscr{F}$ is considered, and $L_1$ and $L_2$ are RALs. 
This is true because for this semiring,
the number of inner-spheres for each RAL $L$ is finite; 
this number is bounded by the number of transitions in an 
annotated automaton for $L$.

Regarding the $\mathscr{T}$ and $\mathscr{N}$ semirings, 
the number of spheres is finite, if and only if, the languages 
are {\em bounded}, that is, 
there is bound or limit on the weight each word can have. 
Fortunately, the boundedness for RALs over $\mathscr{T}$ and $\mathscr{N}$
is decidable. 

For a RAL $L$ over $\mathscr{T}$,
determining whether there exists a bound 
coincides 
with deciding the ``limitedness'' problem for ``distance automata''. 
The later problem is widely known and positively solved in the literature 
   (cf. for example \cite{Has82,Leu91,Sim94,Has00}).
The best algorithm 
is by \cite{Leu91},
and it runs in exponential time in the size of an AL recognizing $L$. 
If $L$ is bounded, then the bound is $2^{4n^3+n\lg(n+2)+n}$, 
where $n$ is the number of states in an AL recognizing $L$.

For a RAL $L$ over $\mathscr{N}$,
determining whether there exists a bound 
is again decidable \cite{WS91}. This can be done in polynomial time. 
However, if $L$ is bounded, the bound is $2^{n\lg n + 2.0566n}$, 
where $n$ is the number of states in an AL recognizing~$L$.

\section{Computing Spheres}
\label{compspheres}

\subsection{Tropical Semiring}

In this section we present an algorithm,
which for any given number $k \in \mathbb{N}$ 
constructs the $k$-th inner-sphere $L^{k}$ 
of a RAL $L$. 

For this, we build a mask automaton ${\cal M}_k$ on the alphabet 
$K=\{0,1,\ldots,k\}$, which formally is as follows:
${\cal M}_k = (P_k, K, \tau_k, p_0, F_k)$, 
where
$P_k=F_k=\{p_0, p_1,$ $\ldots,$ $p_k\}$, and 
\begin{eqnarray*}
\tau_k &=& \{(p_i,n,p_{i+n}) : 0\leq i\leq k, \mbox{ and } 0\leq n\leq k-i \}. 
\end{eqnarray*}

\begin{figure}
\centering
\includegraphics{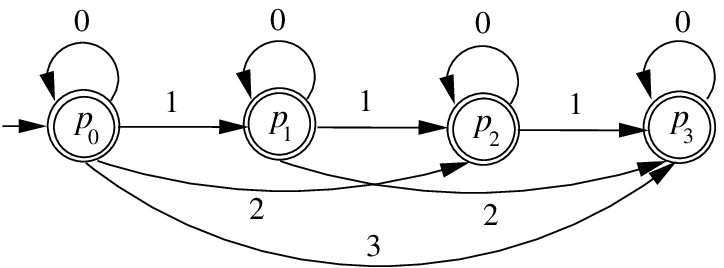}
\caption{Automaton ${\cal M}_3$}
\label{example_fig}
\end{figure}

As an example, we give ${\cal M}_3$ in Fig.~\ref{example_fig}.
The automaton ${\cal M}_k$ has a nice property. 
It captures all the possible
paths (unlabeled with respect to $\Delta$) 
with weight equal to $k$. 
\input epsf           
It can be shown that  
\begin{theorem}
${\cal M}_k$ contains {\em all} the possible paths $\pi$ with $\mbox{\sl weight}(\pi) \leq k$, 
and it does not contain any path with weight greater than $k$.  
\end{theorem}
 
It can be easily seen that the size of automaton ${\cal M}_k$
is ${\cal O}(k^2)$. 
Now by using ${\cal M}_k$, 
we can extract from an annotated automaton $\cal A$ for $L$
all the transition paths with a weight less or equal to $k$,
giving so an effective procedure for computing the $k$-th sphere $L^{(k)}$.

For this, let 
${\cal A} = (P_{\tinya}, \Delta, \tau_{\tinya}, q_0, F_{\tinya})$ 
be an annotated automaton for $L$. 
We construct a Cartesian product automaton 
$$
{\cal C}_k = {\cal A} \times {\cal M}_k = 
(P_{\tinya}\times P_k, \Delta, \tau, (q_0,p_0), F_{\tinya}\times F_k),
$$ 
where 
$
\tau = \{((q,p), r, n, (q',p')) : (q, r, n, q') \in \tau_{\tinya} 
\mbox { and } (p, n, p') \in \tau_k\}.
$
It can be verified that
\begin{theorem}
$\lsem {\cal C}_k \rsem = L^{k}$.
\end{theorem}

\subsection{Fuzzy Semiring}

In order to compute $L^k$, 
where $L$ is a RAL, and $k\in \mathbb{N}$, 
we simply build an annotated automaton $\cal A$
for $L$, and then throw out all the transitions 
weighted by more than $k$. 
Let ${\cal A}'$ be the annotated automaton thus obtained. 
It can be verified that 
\begin{theorem}
$\lsem A' \rsem = L^{k}$.
\end{theorem}

\subsection{Multiplicity Semiring}

One can indeed derive a method for computing inner or outer spheres
for languages over $\mathscr{N}$ using  
complex results spread out in several chapters 
of \cite{Eil76} and \cite{Sak09}. 
We will follow here instead a different, much simpler approach 
based on ordinary automata. 
This approach computes outer spheres.

Let 
${\cal A} = (P_{\tinya}, \Delta, \tau_{\tinya}, p_{{\tinya},0}, F_{\tinya})$ 
be a weighted automaton for $L$. 
From ${\cal A}$ we obtain an ``ordinary automaton''
${\cal B} = (P_{\tinyb}, \Delta, \tau_{\tinyb}, p_{{\tinyb},0}, F_{\tinyb})$,
where $P_{\tinyb} = P_{\tinya}$, $p_{{\tinyb},0}=p_{{\tinyb},0}$, 
$F_{\tinyb}=F_{\tinya}$, and
\begin{eqnarray*}
\tau_{\tinyb} &=& \{\underbrace{(p,r,p'),\ldots,(p,r,p')}_{n} : (p,r,p',n)\in \tau_{\tinyb}\}.
\end{eqnarray*}

We can show that
\begin{theorem}
$(w,k)\in [{\cal A}]$, if and only if, $\cal B$ has $k$ accepting transition paths
that spell $w$.
\end{theorem}

The importance of this theorem is that we now transformed the problem of computing 
inner or outer spheres of $L$ into the problem of computing the sets of words 
spelled out in $\cal B$ by a number of accepting transition paths, which 
is greater or smaller than the sphere index.
For simplicity, we will focus here on outer-spheres.
 
Interestingly, the set of all the words spelled out by at least $k$
accepting transition paths in ${\cal A}$ is indeed computable. 
For this, we present a simple construction
which was hidden as an auxiliary construction in \cite{SH85}. 

The construction is as follows.
Let $\Psi_k$ be the set of $k\times k$ Boolean matrices.   
We build a Cartesian product automaton
\begin{eqnarray*}
{\cal B}^k &=& (P_{\tinyb}^k, \Delta, \tau_{\tinyb}^k, p_{{\tinyb},0}^k, F_{\tinyb}^k)
\end{eqnarray*} 
where
\begin{eqnarray*}
P_{\tinyb}^k &=& \underbrace{P_{\tinyb} \times \ldots \times P_{\tinyb}}_{k} \times \Psi_k\\
p_{{\tinyb},0} &=& (p_{{\tinyb},0},\ldots,p_{{\tinyb},0},\psi), 
								\mbox{ where } \psi[i,j]=0 \mbox{ for } i\neq j, \mbox{ and } \psi[i,j]=1 
								\mbox{ for } i=j\\
\tau_{\tinyb}^k &=& 
\{((p_1,\ldots,p_k,\psi), r, 
	 (p_1',\ldots,p_k',\psi')): (p_i,r,p_i')\in\tau_{\tinyb}, \mbox{ for } i\in[1,k], \mbox{ and } \\
	 & &  
	 \;\;\;\;\;\;\;\;\;\;\;\;
	 \;\;\;\;\;\;\;\;\;\;\;\;
	 \;\;\;\;\;\;\;\;\;\;\;\;
	 \;\;\;\;\;\;\;\;\;\;\;\;\;\;\;\;\;\;\;
	  \psi'[i,j]=1 \mbox{ if } \psi[i,j]=1 \mbox{ or } s_i\neq s_j\} \\ 
F_{\tinyb}^k &=& \underbrace{F_{\tinyb} \times \ldots \times F_{\tinyb}}_{k} \times \{\psi^*\}, \mbox{ where } \psi^*[i,j]=1, \mbox{ for } i,j\in[1,k].
\end{eqnarray*} 

Let $w$ be a word spelled by $k$ or more transition paths in $\cal B$.
Let $\rho_1,\ldots,\rho_k$ be $k$ of these transition paths. 
In the Cartesian product automaton ${\cal B}^k$ we will have a transition path $\rho$
that corresponds to the combination of $\rho_1,\ldots,\rho_k$.   
It can be verified that the last state of $\rho$ in ${\cal B}^k$
will have its matrix equal to $\psi^*$. 
This is because since $\rho_i$ is different from $\rho_j$, 
for each $i,j\in[1,k]$, at some point the matrix of some state in $\rho$
will have 1 for its $i,j$ entry. Then for each subsequent state in $\rho$, 
the correponding matrix will retain 1 in its $i,j$ entry. 
Therefore, if $\rho_1,\ldots,\rho_k$ are accepting transition paths, 
then ${\cal B}^k$ will accept $w$. 
Considering ${\cal B}^k$ From all the above we have
\begin{theorem}
A word $w$ is accepted by ${\cal B}^k$, if and only if, 
there are at least $k$ accepting paths spelling $w$ in $\cal B$.
\end{theorem}

From this theorem and Theorem~\ref{containment}, we then have
\begin{theorem}
$\lfloor L^{\breve{k}} \rfloor= L({\cal B}^k)$.
\end{theorem}

\bibliographystyle{abbrv}
\bibliography{references}{}

\end{document}